\documentclass[showpacs,twocolumn]{revtex4}
\usepackage{graphicx}

\begin{document}
\title{
Positional Order and Diffusion Processes in Particle Systems
}

\author{Hiroshi Watanabe,$^{1}$\footnote{E-mail: hwatanabe@is.nagoya-u.ac.jp}
Satoshi Yukawa,$^{2}$ and Nobuyasu Ito$^3$}

\affiliation{
$^1$Department of Complex
Systems Science, Graduate School of Information Science,
Nagoya University, Furouchou, Chikusa-ku, Nagoya 464-8601, Japan}

\affiliation{
$^2$ Department of Earth and Space Science, Graduate School of Science,  
Osaka University, 1-5 Yamadaoka, Suita, Osaka 565-0871, Japan}

\affiliation{
$^3$ Department of Applied Physics, School of Engineering,
The University of Tokyo, Hongo, Bunkyo-ku, Tokyo 113-8656, Japan
}

\begin{abstract}
Nonequilibrium behaviors of positional order are discussed based on diffusion processes
in particle systems. With the cumulant expansion method up to the second order,
we obtain a relation between the positional order parameter $\Psi$ and
the mean square displacement $M$ to be $\Psi \sim \exp(- {\bf K}^2 M /2d)$
with a reciprocal vector ${\bf K}$ and the dimension of the system $d$.
On the basis of the relation, the behavior of positional order 
is predicted to be $\Psi \sim \exp(-{\bf K}^2Dt)$ when the system involves normal
diffusion with a diffusion constant $D$.
We also find that a diffusion process with swapping positions of particles
contributes to higher orders of the cumulants.
The swapping diffusion allows particle to diffuse without destroying the positional
order while the normal diffusion destroys it.
\end{abstract}


\maketitle

\renewcommand{\v}[1]{{\bf #1}}
\newcommand{\diff}{{\mathrm d}}
\newcommand{\ave}[1]{\left< #1 \right>}
\newcommand{\rhoc}{\rho_{\mbox{\small cp}}}

The melting behavior of the hard-disk system
was reported first by Alder {\it et al.}~\cite{Alder}, 
and they showed that only repulsive interactions can involve
the melting transition.
This melting transition also confirmed in three-dimensional systems
and is now often referred to Alder transition.
However, Mermin ruled out the positional long range order in two-dimensional
particle systems~\cite{Mermin}. Therefore, the melting processes
of two-dimensional systems are different from that of three-dimensional systems.
Halperin, Nelson, and Young proposed the two-dimensional melting theory~\cite{HNY}
based on Kosterlitz Thouless transition~\cite{KT}, and 
Chui proposed another theory predicting the first order transition
based on the grain boundaries excitation~\cite{Chui}.
While many researchers have been studying this problem~\cite{Zollweg, LeeStrandburg, Weber, Fernandez, Jaster,Sengupta},
the nature of two-dimensional melting has been still a matter of debate~\cite{Strandburg}.
So far, most of numerical works have focused on the equilibrium
state of the system mainly using Monte Carlo methods.
Recently, the nonequilibrium behaviors of the bond-orientational order
parameters has been studied to obtain the equilibrium properties of the 
hard-disk systems~\cite{Watanabe}.
These studies are based on a new strategy for the simulation,
called Non-equilibrium relaxation (NER) method~\cite{NERReview}.
Zahn and Maret studied time dependent parameters in two-dimensional
colloidal particle systems~\cite{Zahn}. They pointed out that
static properties are not appropriate measure
to distinguish between the solid and the fluid, 
since the mean square displacement diverges very slowly.
Therefore, it is necessarily to study the dynamic behaviors of 
order parameters in the particle systems.
In this letter, we study the dynamics of the positional order parameter
based on diffusion processes.
We also treat two- and three-dimensional systems at the same time,
since many studies have focused only on the two-dimensional melting and,
to our knowledge, there are less studies about the three-dimensional positional order.

\begin{figure}[tb]
\includegraphics[width=1.0\linewidth]{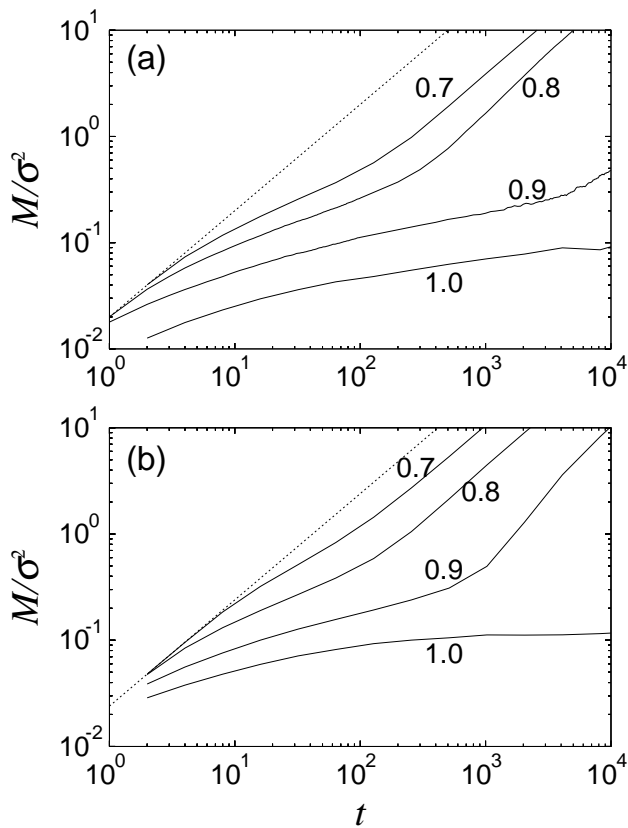}
\caption{
Time evolution of the mean square displacement $M\equiv \ave{u_i^2}$
in (a) two- and (b) three-dimensional systems. The decimal logarithm are taken for the both axes.
The dashed lines denote the diffusion in the low density limit~\cite{low_density_limit}.
Number of particles $N=23288$ for two-
and $N=36000$ for three-dimensional system with the periodic boundary condition.
}
\label{fig_diffusion}
\end{figure}

\begin{figure}[tb]
\includegraphics[width=1.0\linewidth]{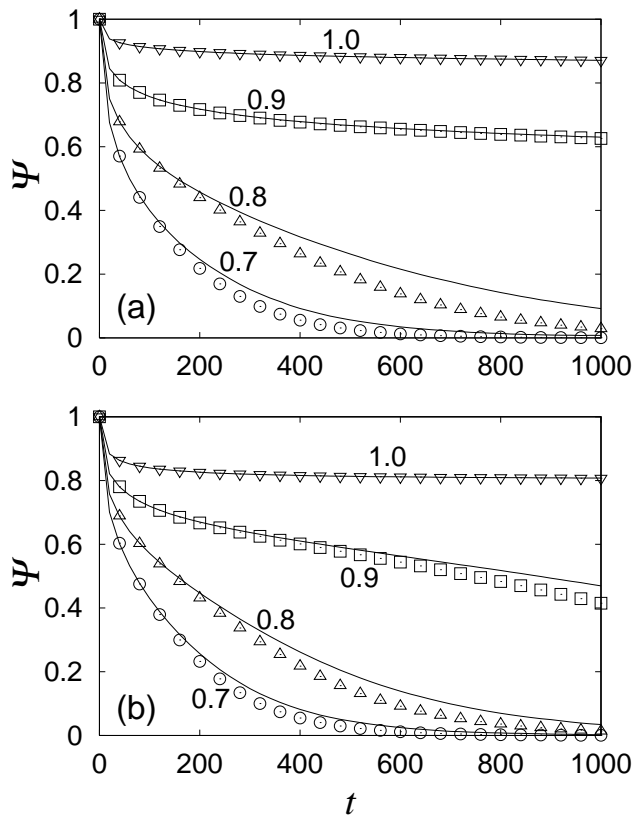}
\caption{
Time evolution of the positional order parameter and
calculated values from the diffusion for
(a) two- and (b) three-dimensional systems.
The solid lines are the positional order parameters 
and the symbols are the calculated values using Eq.~(\ref{eq_diff2pos}).
It shows good agreement in the region where the positional order parameters
are not so small.
}
\label{fig_pos}
\end{figure}


Consider a $d$-dimensional system with $N$ particles.
A positional order parameter $\Psi$ of the system is defined to be
\begin{equation}
\Psi = \frac{1}{N} \sum_j^N \exp (-i \v{K}\cdot \v{r}_j), \label{eq_defpos}
\end{equation}
with the position of the particles $\v{r}_i$ and 
one of the reciprocal vectors $\v{K}$ of the system.
Let $\v{R}_i$ be the equilibrium position of the particle $i$
and $\v{u}_i$ the deviations from it, namely, 
$\v{r}_i = \v{R}_i + \v{u}_i$.
The positional parameter is reduced to be,
\begin{equation}
\Psi = \ave{\exp{\left( -i \v{K}\cdot \v{u}_j \right)}},  \label{eq_psi1}
\end{equation}
where $\ave{\cdots}$ means the average for all particles.
Assuming that the all components of $\v{u}_i$ have the
Gaussian distribution, Eq.~(\ref{eq_psi1}) is reduced to be~\cite{Jancovici},
\begin{equation}
\Psi = \exp{(-1/2 \ave{(\v{K}\cdot\v{u}_i)^2})}. \label{eq_psi2}
\end{equation}
Assuming that $\v{u}_i$ is isotropic, we have,
\begin{equation}
\ave{ \right(\v{K} \cdot \v{u}_i \left)^2 } = K^2 \ave{\v{u}_i^2}/d \qquad (K \equiv |\v{K}|). \label{eq_isotropic}
\end{equation}
From Eqs.~(\ref{eq_psi2}) and (\ref{eq_isotropic}), we obtain
the relation between the positional order and the diffusion to be,
positional order parameter to be,
\begin{equation}
\Psi = \exp \left( -  \frac{K^2 \ave{\v{u}_i^2}}{2d} \right), \label{eq_diff2pos}
\end{equation}
or equivalently,
\begin{equation}
\ave{\v{u}_i^2} = - \frac{2d}{K^2} \ln \Psi. \label{eq_pos2diff}
\end{equation}
Note that, the above argument is the cumulant expansion.
The positional order parameter is the characteristic function of 
displacements. Assuming the distribution of the displacement
to be the Gaussian distribution, we can express the positional
order parameter only with the second order cumulant, which is diffusion.

When a system involves the normal diffusion, the asynptotic behavior 
of the mean square displacement is expected to be,
\begin{equation}
\ave{\v{u}_i^2} \sim 2 d D t, \label{eq_diffusion}
\end{equation}
with a diffusion constant $D$.
From Eqs.~(\ref{eq_diff2pos}) and (\ref{eq_diffusion}), the
asymptotic behavior of the positional order to be,
\begin{equation}
\Psi(t) \sim \exp({-K^2 D t}),
\end{equation}
regardless of the dimension. It implies that 
when the system involves the normal diffusion, the positional order
should decay exponentially with the decay time $D^{-1}$.
This limits the diffusion behavior in solid phases.
In the solid phase of the system with $d\ge 3$,
the parameter $\Psi$ has non-zero value in the equilibrium state.
Therefore, the mean displacement cannot become larger than 
some constant value.
The behavior in two-dimensional solid is different from those in $d \ge 3$.
On the basis of the Halperin-Nelson-Young theory~\cite{HNY},
the positional order parameter in two-dimensional solid behaves as,
\begin{equation}
\Psi(t) \sim t^{-\lambda}. \label{eq_psi2d}
\end{equation}
The mean square displacement in two-dimensional solid behaves
logarithmically as,
\begin{equation}
\ave{\v{u}_i^2} = \frac{4\lambda}{K^2 z} \ln t.
\end{equation}
Therefore, two-dimensional solid cannot involve the normal diffusion process
in the usual sense.


The above arguments are based on the cumulant expansion up to the second order.
In order to check the validity of our arguments, we perform numerical simulations.
For the simplicity, we treat the hard-particle systems.
There are two kinds of methods to study time evolution of a particle system,
a molecular dynamics (MD) method and a Monte Carlo (MC) method.
The MD simulation is performed by integrating
the classical equations of motion. In the hard particle system,
the time evolution is performed by proceeding collision events.
This algorithm is called the Event-Driven method,
which is very efficient to treat the hard-particle system~\cite{Rapaport,Lubachevsky, Isobe}.
When the time evolution of the system is performed by MD,
however, the positional order has oscillations because of the momentum 
conservation. This oscillation prevents us from studying the order parameter,
therefore, MC simulations are performed in this study.

Each system contains $N$ particles with the radius $\sigma$.
The density is normalized to be $\rho = 1$ when the system is in the perfect
square/cubic lattice, that is, $\rho \equiv (2d/L)^dN$ with the dimension of the system $d$
and the linear size of the system $L$.
Throughout this study, the number of particles $N=23288$ for two-
and $N=32000$ for three-dimensional systems and
up to 512 independent samples are averaged for each density.
The step length is set to be $\sigma_s = 0.2 \sigma$ with the radius $\sigma$~\cite{sigma_s}.
At the beginning of each run, the particles are set up 
in the perfect ordered configuration, namely, the hexagonal 
lattice in the two-dimensional and the face centered cubic lattice.
The periodic boundary conditions are taken along the all axes.
The densities from $\rho = 0.7$ to  $1.0$ are studied.


The time evolutions of the mean square displacements $M$ are shown in Fig.~\ref{fig_diffusion}.
We can see the normal diffusions starting after some initial relaxations, for example,
the data of three-dimensional system with $\rho=0.9$ has a bend at $t \sim 10^3$. It implies that 
the normal diffusion started after the positional order are almost destroyed.
The time evolutions of the positional order parameters are shown in Fig.~\ref{fig_pos}.
While the positional order is well approximated by the second order 
cumulant in the region where the positional order parameter is not so small,
there are differences especially in the low densities.

\begin{figure}[tb]
\begin{center}
\includegraphics[width=1.0\linewidth]{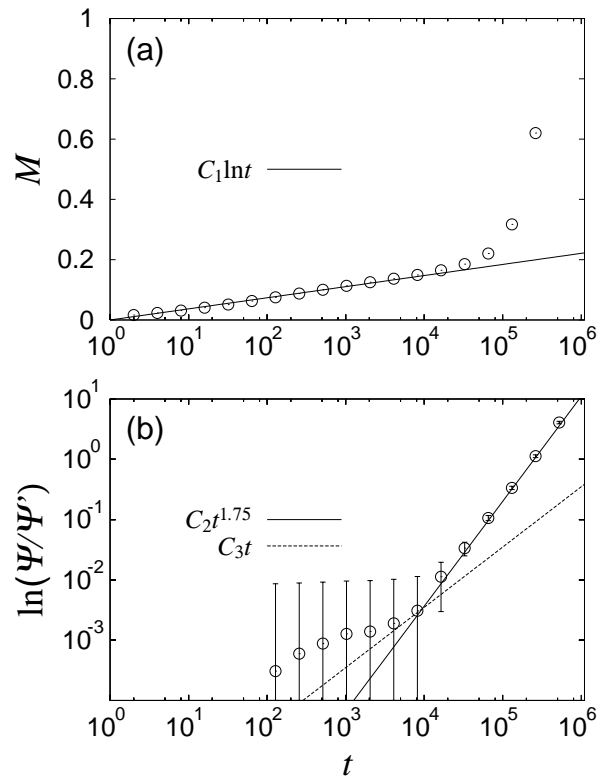}
\end{center}
\caption{
(a) Mean square displacement $M$ of the two-dimensional solid.
The number of particles $N=23288$ and the density $\rho = 0.92$.
The solid line is $C_1 \ln t$ with $C_1=1.6 \cdot 10^{-2}$.
Error bars are smaller than the size of the symbols.
It shows that the crossover from the normal diffusion (logarithmic)
to the swapping diffusion (power-law).\\
(b) The time evolution of the value $\ln(\Psi/\Psi')$, where
$\Psi$ is the positional order parameter with the definition in Eq.~(\ref{eq_defpos})
and $\Psi'$ is the value calculated from diffusion using Eq.~(\ref{eq_diff2pos}).
The decimal logarithms are taken for the both axes.
The solid and dashed lines are drawn for for the guides to the eyes;
The solid line is $C_2 t^{1.75}$ and the dashed line is $C_3 t$
with $C_2 = 4.0 \cdot 10^{-10}$ and $C_3 = 3.5 \cdot 10^{-7}$.
It shows that the exchanging rate increases as $E_r \sim t^{0.75}$.
}
\label{fig_long2d}
\end{figure}

\begin{figure}[tb]
\begin{center}
\includegraphics[width=0.8\linewidth]{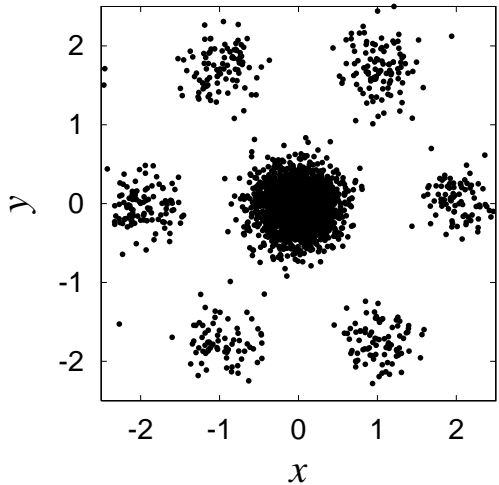}
\end{center}
\caption{
Distribution of displacements $\v{u}_i$ at $t=5\cdot 10^5$ of the two-dimensional system.
A small system with $N=2900$ is shown for visibility.
The density of the system is $\rho=0.92$.
The lattice constant is $a = 2$ and the radius of particles is $\sigma = 0.89$ 
in this scale.
The center $(0,0)$ corresponds to the initial position $\v{R}_i$.
The points around at the center correspond to the normal diffusion
and the six small groups around the center group correspond to the swapping diffusion.
}
\label{fig_distribution}
\end{figure}


These differences are caused by the higher order 
cumulants which are ignored in Eq.~(\ref{eq_psi2}).
The contribution from the higher order cumulants can be explained by a
swapping diffusion process.
In particle systems, there are two kinds of ways to diffuse; by 
normal diffusion and the swapping. While the normal diffusion 
destroys the positional order parameters as described in Eq.~(\ref{eq_diff2pos}),
the swapping does not.
In Fig.~\ref{fig_long2d}, the diffusion behavior in two-dimensional solid is shown.
The density is $\rho = 0.92$ which is high enough than the melting points $\rho_m$,
{\it i.e.}, $\rho_m \sim 0.902$ reported by 
Zollweg and Chester~\cite{Zollweg} 
and $\rho_m < 0.905$ by Weber {\it et al.}~\cite{Weber},
and $\rho_m \sim 0.893$ by Watanabe {\it et al.}~\cite{Watanabe}.
The diffusion shows logarithmic behavior up to $t \sim 10^{4}$ as Mermin predicted~\cite{Mermin}.
However, it varies from the logarithmic behavior
because of the swapping diffusion around at $t \sim 10^{5}$.
The distribution of the displacement $\v{u}_i$ at this time is shown in Fig.~\ref{fig_distribution}.
The points around at the center correspond to the results of the normal diffusion
and the six groups around the center group correspond to that of the swapping diffusion.


In order to treat the effect of the swapping, we consider the system
with two types of diffusion, the continuous diffusion and 
the swapping diffusion with a swapping rate $E_r$
on the lattice with a lattice constant $a$.
The rate $E_r$ denotes the probability to jump to the nearest position
at equilibrium per unit time.
The diffusion with swapping $\ave{\v{u}_i^2}'$ is expressed to be,
\begin{equation}
\ave{\v{u}_i^2}' = \ave{\v{u}_i^2} + d a^2 E_r t,
\end{equation}
with the diffusion without swapping $\ave{\v{u}_i^2}$~\cite{diff_lattice}.
In the following, the positional order parameter 
calculated from Eq.~(\ref{eq_diff2pos}) is denoted by $\Psi'$ in order to
distinguish from the original definition in Eq.~(\ref{eq_defpos}).
Using $E_r$, $\Psi'$ can be expressed to be,
\begin{eqnarray}
\Psi' &=& \exp \left( -  \frac{K^2 \ave{\v{u}_i^2}'}{2d} \right) \\ \nonumber
&=& \Psi \exp{( -2\pi^2 E_r t)}. \label{eq_psi_psip}
\end{eqnarray}
Therefore, $\Psi'$ is always smaller than $\Psi$, since $E_r > 0$.
The contribution of the higher order cumulants is expressed to be,
\begin{equation}
\ln \left( \Psi/\Psi' \right) = 2 \pi^2 E_r t. \label{eq_psip_log}
\end{equation}
The time evolution of the value $\ln \left( \Psi/\Psi' \right)$ is shown
in Fig.~\ref{fig_long2d}(b).
It increases as $\sim t^{1.75}$, which is faster than linear increase.
If the exchanging rate $E_r$ is constant, the value should increase as $\sim t$.
Therefore, the exchanging rate $E_r$ increases. It implies that
the destruction of the positional order enhances the swapping of the particles.


To summarize, we study the dynamics of the 
positional order in the particle systems
based on the diffusion processes.
We discuss the relation between the positional
order parameter $\Psi$ and the mean square displacement $M$ with the
cumulant expansion.
We find that there are two kinds of diffusion processes in particle systems,
one is the normal diffusion and another is the swapping diffusion
which allows particles to diffuse without destroying the positional order.
These diffusion processes can be understand as cumulants of the displacements;
the normal diffusion is the second order cumulant, and
the swapping diffusion contributes to the higher orders.
This swapping diffusion process will play important roles
in systems with two or more kinds of particles in high density region.
The presented arguments are very general, 
and applicable to other systems with general pair potentials.
Studying dynamic aspects of Alder transitions based on the cumulant expansion 
should be a further issue.

We thank M. Mori for helpful suggestions.
The computation was partially carried out
using the facilities of the Supercomputer Center,
Institute for Solid State Physics, University of Tokyo.
This work was supported by the 21st  COE program,
``Frontiers of Computational Science", Nagoya University and
the Grant-in-Aid for Scientific Research (C), No.~15607003,
of Japan Society for the Promotion of Science.


\begin{thebibliography}{99}
\bibitem{low_density_limit}In the low density limit, the mean square displacement $M$
of MC is exactly obtained to be $M = \sigma_s^2 d /(d+2) \cdot t$ with
the step length $\sigma_s$ and the dimension of the system $d$.
Since the step length is set to be $\sigma_s = 0.2 \sigma$ in this study,
the lines $0.04 d/(d+2) \cdot t$ are shown in Fig.~\ref{fig_diffusion}.
\bibitem{sigma_s} We have also performed simulations with other values of step length
$\sigma_s = 0.01, 0.05$ and $0.1$,
and confirmed that the parameter only changes the time scale of the system and
the results are not changed qualitatively.
\bibitem{diff_lattice} Consider a diffusion process on the grid with 
the lattice constant $a$. If particles jump randomly to the next site
every $\tau$, the system involves diffusion with a diffusion constant
$D = a^2/2\tau$ and the exchanging rate $E_r$ is denoted by $E_r = \tau^{-1}$.
\end{thebibliography}
\end{document}